\documentclass[a4paper,12pt]{article}
\author{Michal Zaja\v{c}ek}
\usepackage[utf8]{inputenc}
\usepackage{graphicx}
\usepackage{amsthm}
\usepackage{amssymb}
\usepackage{epstopdf}
\usepackage{url}
\usepackage{amsthm}
\usepackage{natbib}
\usepackage{amsmath}
\usepackage{wasysym}
\usepackage{anysize}
\marginsize{3cm}{3cm}{1cm}{1cm}

\usepackage[ps2pdf,unicode]{hyperref}   
\hypersetup{pdftitle=}
\hypersetup{pdfauthor=Michal Zaja\v{c}ek}
\title{\textbf{Polarimetry narrows down the possibilities for the Dusty S-cluster Object (DSO/G2) in the Galactic centre}}
\date{}
\author{Michal Zaja\v{c}ek$^{*}$, Andreas Eckart, and Banafsheh Shahzamanian}

\newcommand{\apj}{ApJ}
\newcommand{\apjl}{ApJL}

\newcommand{\mnras}{MNRAS}

\newcommand{\nat}{Nature}
\newcommand{\aap}{A\& A}

\begin{document}

\maketitle

\begin{center}
University of Cologne, Zülpicher Strasse 77, 50937 Cologne, Germany\\
$^{*}$\url{zajacek@ph1.uni-koeln.de}\\
\vspace*{0.5cm}
Accepted by the Editors of \textit{The Observatory (A Review of Astronomy)}
\end{center}

\begin{abstract}

There have been many speculations about the character of the dusty object moving fast in the vicinity of the Galactic centre black hole. The recent detection of polarized continuum emission provides new constraints for the models. The fact that the object is intrinsically polarized implies that it is non-spherical. The authors propose that a young star developing a bow shock can explain the main characteristics. However, more observations in the future are needed for the final confirmation of the nature of the source.

\end{abstract}

The object known as the Dusty S cluster object (DSO; where S cluster is the name of the innermost stellar cluster in the Galactic centre), often also denoted as G2, was shortly after its discovery \citep{1} considered as a small gas and dust cloud of only three Earth masses. However, subsequent monitoring by the Very Large Telescope of the European Southern Observatory as well as Keck telescopes has shown that the object does not tidally stretch in a way as we would expect for a simple core-less gas cloud. Instead, the dusty source has stayed more compact than expected and survived the closest passage to the black hole intact in the spring of 2014 \citep{2,3}. 

Despite the compactness, the dispute about the character of the object has continued, mainly because it has been not possible to directly resolve the internal structure of the enigmatic object as it is possible, for example, for nearby young stars in the Orion star-forming region. To make things more complicated, some experts claim \citep{1,4} that they do detect the tidal stretching of the source. As a result, many scenarios for the nature of the DSO have been proposed. They can be mostly grouped into three categories: a core-less cloud \citep{1} or streamer \citep{4,5}, a dust-enshrouded star \citep{6,7}, and a binary scenario – either binary merger \citep{3,8} or disruption \citep{7} of both components, where one of them can escape the Milky Way entirely as a so-called hypervelocity star.

However, there is a way to go partially around the angular resolution problem. One can try to study the polarization properties of the incoming electromagnetic signal to see if the source as a whole is polarized or not. Polarized sources have a preferred plane in which the electric field vectors are oscillating, which gives hints about their internal geometry as well as radiative processes.

It was quite a surprise when it was discovered \citep{9} that the DSO is an intrinsically polarized source in the near-infrared $K_{\rm s}$-band (2.2 micrometers). Whereas surrounding stars close to its position have a degree of polarization close to zero, and hence their emission is not polarized, the DSO exhibits a polarization degree of around 30 percent for four consecutive epochs (2008, 2009, 2011, and 2012). This implies that the source must deviate from a spherical symmetry, otherwise the individual polarization contributions would cancel out.

The detection of polarized emission adds a new constraint on the character of the object. In general, the DSO is a very faint source in an extremely crowded stellar field – the number density of stars in the central few light years is about 10 million times that in the Sun's neighborhood. Therefore disentangling the emission of the DSO from the surrounding sources is often challenging. In addition, it is not possible to resolve the brightness distribution as for nearby objects. As a consequence,  it is necessary to carefully combine orbital dynamics, spectral properties, and at last the  polarimetry to see the full picture of the mosaic.

In the paper published in \textit{Astronomy \& Astrophysics}, authors \citep{9} also construct a numerical radiative transfer model of the DSO. The model consists of typical ingredients of young stars: a star at the centre of the DSO, which is the source of thermal photons, is surrounded by a dusty envelope and bipolar cavities due to outflows, which together reprocess the emission of the star – UV and optical photons are absorbed by dust particles and reemitted at longer wavelengths, mostly in the near- and the mid-infrared domains. Furthermore, the photons emitted by the star and the dust are scattered by dust particles, which is the source of polarized emission in the model.

 Moreover, since the DSO is expected to move supersonically close to the black hole, a bow shock is formed ahead of the star \citep{10}. All of these components, which one would expect for a supersonic young star in the Galactic centre region, lead to a significant non-spherical nature of the source, which gives rise to the overall polarized near-infrared emission. Not only is the model successful in explaining the polarization properties, it can also match other observed characteristics of the DSO, namely a significant near-infrared excess or “reddening” due to dust emission and broad emission hydrogen lines, which arise due to the Doppler broadening either because of the material flowing in towards the star (accretion) or by gas outflows or winds, which are both a typical feature of young stars \citep{2}.

It could be argued that the overall non-spherical shape is caused by the gradual prolongation of the gaseous component by tidal forces rather than the model above. However, the DSO/G2 source does not show convincing signs of tidal interaction in both line and continuum emission \citep{2,3}. Tidal stretching would be expected for a core-less cloud or a star with an extended envelope with a length-scale of about 100 AU. In that case the source would be tidally stretched along the orbit by a factor of a few \citep{2}, which was not detected during the peribothron passage \citep{2,3}, when the effects of the orbital foreshortening are minimized. In fact, the DSO is fully consistent with being a point source \citep{3}. 

Therefore, based on the compactness and a prominent IR excess, a pre-main-sequence star surrounded by a non-spherical dusty envelope (envelope with bipolar cavities) seems to be a more natural scheme to explain the continuum and line emission characteristics. In the framework of this scenario, a bow shock forms due to an expected supersonic motion close to the pericentre, which further breaks the spherical symmetry.

Further monitoring of the source will help to test the proposed model, mainly by the means of orbital dynamics. In case the source will not deviate from a simple Keplerian ellipse, it must be a compact object, not a cloud. On the other hand, the core-less cloud would sooner or later start spiraling in towards the black hole because of the interaction with the surrounding ambient medium.

It remains a small puzzle though how such a young star as proposed to explain the DSO phenomenon can be formed and subsequently orbit so close to the black hole for a longer period of time – possibly several hundred thousand years, which is the estimated age of class 0 and class 1 protostellar objects \citep{2}. Thanks to the computer  modeling, this problem can be partially tested by means of numerical experiments.  It was already confirmed \citep{11} that in-situ star formation close to the black hole can take place when a cold molecular cloud of about 100 Solar masses falls in towards the black hole from the region where a molecular circum-nuclear disc is located that contains clumps of a similar mass (approximately 1.5 – 6 parsecs from Sgr A*). In this model, the critical density for the onset of the collapse is reached by the tidal focusing thanks to the black hole's gravity – one can talk about the black hole assisted star formation. Another proposed scenario is the gravitational instability and the fragmentation of a massive accretion disc encircling the black hole \citep{12}, which is supported by the observed stellar disc containing massive young stars having an age of only a few million years. Ongoing star formation in the central 2 parsecs was also supported by recent radio and infrared observations \citep{13}  in terms of finding localized water and SiO masers and identifying infrared excess sources whose spectral energy distribution is consistent with massive young stellar objects.

 Since the  star formation close to the massive black hole has many intricacies, several important details of how stars are formed at the Galactic centre remain still blurred. New powerful instruments in the near future, such as James Webb Space Telescope or  European Extremely Large telescope,  will certainly shed new light on the problem. Regardless of some remaining theoretical problems, the observations seem to show that star formation can proceed in different environments throughout the Galaxy – from the close vicinity of the supermassive black hole at the Galactic centre all the way to the Galaxy outskirts.


\tiny
\nocite{*}


\end{document}